\documentclass[5p,times,authoryear,twocolumn]{elsarticle}

\usepackage{booktabs}
\usepackage{mathtools}
\usepackage{setspace} % for \onehalfspacing and \singlespacing macros
\usepackage{etoolbox}
\usepackage{enumitem}
\usepackage{fancyvrb}
\usepackage{amssymb}    % The amssymb package provides various useful mathematical symbols
\usepackage{subfiles}   % Paper sections in separate files
\usepackage{verbatim,tcolorbox}   % Block comments
\usepackage{ulem}      %crossing out text using \sout{}
\usepackage{pdfpages} %added it to enable pdf display (for the mandate)

\usepackage[dvipsnames]{xcolor}
\definecolor{ao}{rgb}{0.0, 0.5, 0.0}

\usepackage{url}

\usepackage{breakurl}
\usepackage[unicode,breaklinks]{hyperref}
\hypersetup{
	colorlinks=true,
   	linkcolor=blue,
   	citecolor=blue,
    filecolor=blue,
   	urlcolor=blue}

\newcommand{\furl}[1]{\footnote{\url{#1}}}

\usepackage{rotating}

\usepackage{multirow}
\usepackage{tabularx}
\newcolumntype{L}[1]{>{\raggedright\arraybackslash}p{#1}}
\newcolumntype{C}[1]{>{\centering\arraybackslash}p{#1}}
\newcolumntype{R}[1]{>{\raggedleft\arraybackslash}p{#1}}

\renewcommand{\emph}[1]{\textit{#1}}

\begin{document}

\begin{frontmatter}

\title{Hey GPT-OSS, Looks Like You Got It – Now Walk Me Through It! An Assessment of the Reasoning Language Models Chain of Thought Mechanism for Digital Forensics}

\author[unia]{Gaëtan Michelet\corref{cor1}}
\address[unia]{Chair for Cybersecurity, University of Augsburg, Augsburg, Germany}
\cortext[cor1]{Corresponding author.}
\ead{gaetan.michelet@uni-a.de}

\author[unia]{Janine Schneider}
\ead{janine.schneider@uni-a.de}

\author[unia]{Aruna Withanage}
\ead{aruna.withanage@uni-a.de}

\author[unia]{Frank Breitinger\corref{cor1}}
\ead{frank.breitinger@uni-a.de}
\ead[url]{https://www.FBreitinger.de}

\begin{abstract}
The use of large language models in digital forensics has been widely explored. Beyond identifying potential applications, research has also focused on optimizing model performance for forensic tasks through fine-tuning. However, limited result explainability reduces their operational and legal usability.
Recently, a new class of reasoning language models has emerged, designed to handle logic-based tasks through an `internal reasoning' mechanism. Yet, users typically see only the final answer, not the underlying reasoning.
One of these reasoning models is gpt-oss, which can be deployed locally, providing full access to its underlying reasoning process. This article presents the first investigation into the potential of reasoning language models for digital forensics. Four test use cases are examined to assess the usability of the reasoning component in supporting result explainability. The evaluation combines a new quantitative metric with qualitative analysis. Findings show that the reasoning component aids in explaining and validating language model outputs in digital forensics at medium reasoning levels, but this support is often limited, and higher reasoning levels do not enhance response quality.
\end{abstract}


\end{frontmatter}

%%%%%%%%%%%%%
\section{Introduction}\label{sec:introduction}

Since the release of ChatGPT in 2022, large language model (LLM) capabilities have been extensively explored for digital forensics tasks.
For example, LLMs have been evaluated in terms of their ability to write scripts related to digital forensics \citep{Wickramasekara2025auto}, to help with volatility-based memory forensics \citep{oh2024volgpt,lang2025leveraging}, and to generate drafts for different parts of the forensic report \citep{michelet2024chatgpt}.
In addition, recent research has also investigated fine-tuning LLMs for answering general digital forensics questions \citep{sharma2025forensicllm} and chat summarization \citep{michelet2025finetuning}.

However, one of the main problems with LLMs is that it is difficult to comprehend, and thereby to explain, how these models obtain their results, which are then returned to the user.
Consequently, there is a need to better understand LLM-generated responses (LLM explainability), especially if they are intended for digital forensics, as the outputs may have to be explained and justified in court. 
One step towards easy-to-use LLM explainability would be to generate indications of how the model reached an answer, which may simplify the verification process of LLM results.

A possible solution might be the novel Chain of Thought (CoT) mechanism of the newly released Reasoning Language Models (RLMs) by companies such as OpenAI\footnote{\url{www.openai.com}}. 
This new type of LLM adds a step-by-step reasoning layer that helps the model solve complex tasks such as arithmetic or coding tasks. Note that we will use the reasoning process, CoT, and the reasoning part interchangeably for the rest of the paper. 
In summary, these models first generate a step-by-step internal reasoning before generating the final answer, which is displayed to the user.

Current common models are
Deepseek\footnote{\url{www.deepseek.com/}}'s Deepseek-R1 and Qwen\footnote{\url{www.qwen.ai/home}}'s QwQ, which are freely accessible and locally deployable RLMs \footnote{\url{www.huggingface.co/deepseek-ai/DeepSeek-R1}, last accessed 2025-11-25}\footnote{\url{www.huggingface.co/Qwen/QwQ-32B}, last accessed 2025-11-25}. They were released in January 2025 and March 2025, respectively. OpenAI released gpt-oss \footnote{\url{www.huggingface.co/collections/openai/gpt-oss-68911959590a1634ba11c7a4}, last accessed 2025-11-25}, a RLM offered with 20 or 120 billion parameters, in August 2025.
The smaller version is supposed to be usable on a 16GB GPU.

All three models can be deployed and operated locally, making them well-suited for digital forensics investigations by eliminating data sharing with external AI providers and enabling direct access to the model's reasoning output. Once extracted (the reasoning output is usually not shown to a user), this chain of thought component can enhance the comprehensibility, verifiability, and explainability of LLM-generated results, thereby increasing the overall trustworthiness and reliability of the forensic process.
In this paper, we investigate the potential and impact of RLMs for digital forensics.
Our central research question is:

\begin{quote}
    To what extent can the result from the chain of thought mechanism of reasoning language models help to understand and justify the obtained results when undertaking digital forensics tasks?
\end{quote}

To answer this question, we decided to conduct experiments using OpenAI's gpt-oss-20b model.
This choice was determined by the possibility of running the model locally and the easy access to the CoT space.
In addition, gpt-oss is one of the newest published reasoning models. Its balanced size makes it easily deployable.

For this purpose, we created four test scenarios and evaluated both the reasoning process as well as the final answer for each of the scenarios.
To evaluate the results, we decided to develop the following secondary research questions: 
\begin{enumerate}[label=RQ\arabic*]
    \item \hypertarget{rq1}{What is the quality of the CoT generated by gpt-oss for a given digital forensics task?} 
    \item \hypertarget{rq2}{How does the adjustable level of reasoning impact the quality of the CoT?} 
    \item \hypertarget{rq3}{Does the quality of the CoT impact the quality of the final answer?} 
    \item \hypertarget{rq4}{Can the CoT effectively assist in understanding and explaining why the model reached a certain conclusion?} 
\end{enumerate}

\subsection{Contribution}
To the best of our knowledge, we are the first to investigate the potential of RLMs for explainability and justifiability of LLM-generated answers to digital forensics tasks.
Therefore, we develop four unique testing scenarios for RLM testing and evaluation, 
adapt standard RLM evaluation metrics to evaluate the model's CoT quality,
introduce additional metrics to assess the final responses' quality, 
evaluate the usability of OpenAI's gpt-oss model's CoT using  our test scenarios through quantitative metrics and qualitative evaluation techniques, 
and examine the impact of different factors on the quality of the CoT and final response. 

\subsection{Outline}

The remainder of this paper is structured as follows:
We present the background on RLMs and important related work in Sec.~\ref{sec:background}.
This is followed by the description of the methodology in Sec.~\ref{sec:methodology}.
The methodology section includes the description of the four testing scenarios and the testing environment. 
It also contains details on the setup of the experiments, important parameters, and the description of the evaluation process.
After that, we present the results of our study in Sec.~\ref{sec:results}, the discussion of the results in Sec.~\ref{sec:discussion}, and limitations and future work in Sec.~\ref{sec:limitations}.
We conclude our paper in Sec.~\ref{sec:conclusion}.

\subsection{Notice on Reasoning Models}
The reasoning language model used and presented in this paper does not actually think or reason. Similarly to normal language models, they predict one of the most probable next tokens given the current context, and thus in an iterative way (i.e., it generates one token after another, appending the last generated token to the context). This ability to use `reasoning' is due to the training process, which tries to mimic human reasoning.

%%%%%%%%%%%%%%%%%%%%%%%%
\section{Background and Related Work}\label{sec:background}

\subsection{Large Language Models in Digital Forensics}

The use of LLMs for digital forensics tasks is now a fairly well-researched field. The investigation of artifacts left behind by the use of LLMs on a system is out of scope for our study. We will therefore focus on the research already conducted for the application of LLMs to the digital forensics process.
Some of the papers focused on the application of LLMs to a specific task or set of tasks.

For example, \cite{henseler2023chatgpt} investigated the capabilities of large language models in three digital forensics use cases: 
generating queries for Hansken, 
summarizing, evaluating, and visualizing electronic communications, and analyzing search outputs.
\cite{michelet2024chatgpt} examined the potential of large language models such as ChatGPT and Llama-2 to support the writing of forensic reports. 
\cite{Wickramasekara2025auto} introduced AutoDFBench, an automated framework for testing and benchmarking AI-generated digital forensic code, including outputs from LLMs.

LLMs' applicability was also investigated in memory forensics, where \cite{oh2024volgpt} presented volGPT, a novel approach for ransomware triage in memory forensics that combines an LLM with the Volatility framework.  
Along the same lines, \cite{lang2025leveraging} investigated the integration of LLMs into memory forensic workflows to enhance the detection of stealthy malware and advanced persistent threats.

The capabilities, limits, and risks associated with the use of LLMs in digital forensics have also been widely discussed:

\cite{scanlon2023chatgpt} evaluated the capabilities and limitations of ChatGPT, in particular GPT-4, in a range of digital forensic contexts. \cite{scanlon2023digital} discussed the transformative impact of LLMs 
%such as BERT, GPT-3, GPT-4, LLaMA, and others 
in the field of digital forensics, reflecting on how the rapid adoption of such models has sparked promising applications, but also general debates about trust, authenticity, and reliability when using LLMs in various domains.
\cite{dinis-oliveira2023chatgpt} explored how ChatGPT could serve as a virtual assistant for lawyers, judges, and victims to interpret and manage digital forensics expert evidence.

In addition, there is also research on other aspects of LLMs, such as fine-tuning them. \cite{sharma2025forensicllm} presented ForensicLLM, a fine-tuned local LLaMA-3.1–8B model specifically adapted for digital forensics tasks and trained with question and answer pairs extracted from research articles and curated digital artifacts.

Practical recommendations for fine-tuning LLMs for digital forensics tasks were developed by \cite{michelet2025finetuning}.
To demonstrate the applicability of their approach, the authors presented a case study on chat summarization, evaluating the performance of several fine-tuned models.
Finally, \cite{cho2024exploring} explored the use of LLMs and fine-tuning for author profiling in digital text forensics (trying to detect, for example, the gender or age of an author based on the writing style).

More information on the applicability of LLMs in digital forensics can be found in \cite{wickramasekara2024exploring}, who explored the use of LLMs for different aspects of Digital Forensics, and \cite{xu2025large}, who reviewed different integrations of LLMs to the digital forensics process. 
While extensive work has been done on LLMs in digital forensics, none of the papers examine explainability in detail.

\subsection{Reasoning and Large Language Models}
Before the creation of specialized reasoning models, researchers used few-shot prompting and prompt engineering to simulate and obtain the `thought process' of LLMs.

For example, \cite{wei2022chain} investigated how a few-shot prompting combined with a chain of thought can enhance the reasoning capabilities of LLMs.
Their experiments demonstrated that this approach enables LLMs to solve complex tasks in arithmetic, commonsense, and symbolic reasoning more effectively than standard prompts.
Other methods are presented by \cite{plaat2024reasoning}, who reviewed different approaches that leverage LLMs' reasoning capabilities through prompting.

\cite{creswell2022faithful} tried to design a workflow producing faithful multi-step reasoning using LLMs. Their method integrates reasoning steps generated by two fine-tuned LLMs, independently handling the selection and inference processes. The selection model selects facts from the context that might be useful to answer the question and provides them to the inference model. This inference model will then reason using the provided content. The generated text is added to the context, constituting one of the reasoning steps. The process is repeated until the question is answered.

Since the rise of these reasoning process ideas, different surveys have been published, summarizing the state of the art of reasoning and LLMs. For example, 
\cite{patil2025advancing} provided a comprehensive survey focusing on enhancing the reasoning capabilities in LLMs, and  \cite{huang2023reasoning} provided a comprehensive review targeting the LLMs' reasoning capabilities.

\subsection{Reasoning Language Models}

A regular LLM is trained to predict the next token based on the current context.
Its reasoning abilities emerge implicitly, meaning the model can solve problems that require logical steps, but it is not explicitly trained to do so or to display the reasoning process.
RLMs, on the other hand, are specifically trained to produce explicit reasoning steps before providing a final answer.
These intermediate steps are called chain of thoughts.
When an RLM receives a command prompt, it breaks the task into smaller steps and tries to solve them sequentially.
After generating these reasoning steps, the model produces the final answer based on its intermediate reasoning.
Overall, the CoT mechanism helps models perform more accurately on tasks that require reasoning, while also making their internal thought processes more transparent and interpretable.
More details on the RLM components are presented by \cite{besta2025reasoning}, who surveyed RLMs' related work.

While studies usually focus on evaluating the final answer by using benchmark datasets, some work mentions ways to evaluate the reasoning process itself. It is sometimes an element of the survey \citep{patil2025advancing,plaat2024reasoning,huang2023reasoning}, and sometimes the target of the survey \citep{mondorf2024accuracy}.

More recently, \cite{lee2025evaluating} highlighted inconsistencies in the reasoning evaluation process and pointed out the variability of available metrics based on the reasoning task at hand. They therefore propose a structured taxonomy with four criteria that can be applied to any reasoning task: factuality, validity, coherence, and utility. These four criteria will also be utilized within this study.

\subsection{LLM/RLM Specifics}
This section provides some background knowledge on LLMs and RLMs, in addition to the related work, which is essential to understanding our experiment. 

\subsubsection{Chat Templates and Prompts}
Chat templates are pre-defined structures for interacting with LLMs.
They typically specify roles (like `assistant' or `user') and provide contextual instructions.
The main purpose of these templates is to guide the model's behavior, ensuring consistent, coherent, and task-appropriate responses.
Using templates allows users to control tone, style, and output type without modifying the model itself, making interactions more predictable and efficient.
To utilize a chat template to generate effective prompts, the template's pre-defined fields simply have to be populated with task-specific instructions and contextual information.
Fig.~\ref{fig:chat_template_gpt_oss} shows an example of the gpt-oss chat template used in our experiments.

\begin{figure}[t]
\centering
\caption{Chat template for the gpt-oss reasoning language model (simplified and arranged for readability). Green text represents the context submitted to the model (i.e., the system information). ROLE (sometimes referred to as the model identity) and REASONING LEVEL (low, medium, or high) are manually set, while DATE is automatically computed when the template is applied. Text depicted in blue shows the manually created user prompt. The orange text represents the model's inference (i.e., the text that the model will generate).}
\label{fig:chat_template_gpt_oss}
\begin{minipage}{\columnwidth}
{\footnotesize
\begin{tcolorbox}[opacityback=100]
\begin{Verbatim}[commandchars=\\\{\}]
\textcolor{PineGreen}{<|start|>system<|message|>ROLE}
\textcolor{PineGreen}{Knowledge cutoff: 2024-06}
\textcolor{PineGreen}{Current date: DATE}

\textcolor{PineGreen}{Reasoning: REASONING LEVEL}

\textcolor{PineGreen}{# Valid channels: analysis, commentary, final.}
\textcolor{PineGreen}{Channel must be included for every message.}
\textcolor{PineGreen}{<|end|>}

\textcolor{blue}{<|start|>user<|message|>PROMPT}
\textcolor{blue}{<|end|>}

\textcolor{orange}{<|start|>assistant<|channel|>analysis<|message|>CoT}
\textcolor{orange}{<|end|>}

\textcolor{orange}{<|start|>assistant<|channel|>final<|message|>ANSWER}
\textcolor{orange}{<|return|>}
\end{Verbatim}
\end{tcolorbox}
}
\end{minipage}
\end{figure}

\subsubsection{RLM Parameters}

RLMs rely on several parameters that strongly influence the quality and reliability of the reasoning process. For our study, the following parameters are relevant:
\begin{itemize}
    \item The \emph{maximum number of tokens} generated sets an upper bound on the model's output length; if it is too low, reasoning chains may be truncated, whereas overly large values can lead to redundant or unfocused outputs.
    
    \item The \emph{sampling strategy} determines how tokens are selected during text generation.
Deterministic methods produce consistent but rigid results, while probabilistic methods introduce controlled randomness that can enhance reasoning diversity and creativity.

    \item The \emph{temperature parameter} further adjusts this randomness by shaping the probability distribution. Low values lead to deterministic, precise reasoning, while higher values yield more exploratory but potentially inconsistent results.
\end{itemize}

In our experiments, we determined the parameters in advance through pre-testing.
Furthermore, the sampling of the generated tokens and the temperature vary between the different experimental setups as described in Sec.~\ref{sec:experiments}.

%%%%%%%%%%%%%%%%%%%%%%%
\section{Methodology}\label{sec:methodology}

This section presents the methodological framework and experimental design of our study.
In summary, we conducted four experiments comprising multiple runs with varying parameter configurations to systematically analyze gpt-oss behavior under different conditions. In total, we conducted and evaluated 60 experiments.
The following subsections describe the setup, parameter choices, and evaluation procedures in detail, ensuring transparency and reproducibility of the results.

\subsection{Experimental Setup}\label{sec:experiments}
For our study, we simulated four typical tasks:
\begin{itemize}
    \item Suspicious Message Detection (SMD): The model is provided with information about a case investigated and messages extracted from a chat conversation. It must then determine whether the chat is relevant to the investigation or not.
    \item Bash History Analysis (BHA): The model is provided with the bash history extracted from a laptop belonging to someone suspected of trying to attack a server. It must determine if an attack was undertaken from that laptop and, if so, who was the target.
    \item Methodology Generation (MG): The model receives information about a case under investigation and the associated investigative questions. It must then design a methodology that can be used to answer these questions.
    \item Timeline Analysis (TA): The model is provided with information about a suspected data theft and a timeline obtained from a forensic image. It must then determine whether the timeline data can confirm the theft.
\end{itemize}
Further details about the tasks can be found in~\ref{sec:appendix}.
The full prompts can be reviewed on GitHub \footnote{\url{https://github.com/Michelet-Gaetan/Hey-GPT-OSS-Looks-Like-You-Got-It-Now-Walk-Me-Through-It}}.

The tasks were chosen based on three criteria: in practice, they must either occur frequently or be very time-consuming to perform manually, require a certain form of reasoning, and the input data for the task must be in a format supported by gpt-oss.
For each task, a prompt is generated using the gpt-oss chat template with the following values:
\begin{itemize}
    \item ROLE: ``You are a large language model specialized in digital forensics investigation.''
    \item REASONING LEVEL: Low, Medium, and High, depending on the current run.
    \item PROMPT: Designed to focus on the context, the input data, the instructions, and the output format.
\end{itemize}

For each task, one prompt is generated that is used under different conditions (i.e., with various parameters) as can be seen in Table~\ref{tab:experiments}. We have a total of 60 tests.

\begin{table}[h]
\resizebox{\columnwidth}{!}{%
% \begin{tabular}{@{}lll@{}}
\begin{tabular}{l|l|l|l|l|l}
\toprule
% 1, SMD, reasoning, temperature, sampling
\textbf{ID} & \textbf{Task} & \textbf{Reasoning} & \textbf{Temperature} & \textbf{Sampling} & \textbf{Repetitions} \\ \midrule
1-4 & SMD & Low & 0.7 & Yes & 4\\
5 & SMD & Low & 0 & No & 1 \\
6-9 & SMD & Medium & 0.7 & Yes & 4 \\
10 & SMD & Medium & 0 & No & 1 \\
11-14 & SMD & High & 0.7 & Yes & 4 \\
15 & SMD & High & 0 & No & 1 \\
16–19 & BHA & Low & 0.7 & Yes & 4\\
20 & BHA & Low & 0 & No & 1 \\
21–24 & BHA & Medium & 0.7 & Yes & 4 \\
25 & BHA & Medium & 0 & No & 1 \\
26–29 & BHA & High & 0.7 & Yes & 4 \\
30 & BHA & High & 0 & No & 1 \\
31–34 & MG & Low & 0.7 & Yes & 4\\
35 & MG & Low & 0 & No & 1 \\
36–39 & MG & Medium & 0.7 & Yes & 4 \\
40 & MG & Medium & 0 & No & 1 \\
41–44 & MG & High & 0.7 & Yes & 4 \\
45 & MG & High & 0 & No & 1 \\
46–49 & TA & Low & 0.7 & Yes & 4\\
50 & TA & Low & 0 & No & 1 \\
51–54 & TA & Medium & 0.7 & Yes & 4 \\
55 & TA & Medium & 0 & No & 1 \\
56–59 & TA & High & 0.7 & Yes & 4 \\
60 & TA & High & 0 & No & 1 \\ \bottomrule
\end{tabular}%
}
\caption{Details on the different experiments conducted in this study. SMD refers to the Suspicious Message Detection experiments, BHA refers to the Bash History Analysis experiments, MG refers to the Methodology Generation experiments, and TA refers to the Timeline Analysis experiments.}
\label{tab:experiments}
\end{table}

Note that while using probabilistic sampling adds randomness to the generation, not using sampling produces consistent results.
Therefore, generating several texts without sampling is not required, as they will always be identical.
However, generating several texts with sampling enabled allows us to evaluate how the model performs when randomness is included in the generation process.

\subsection{Evaluation}
For each experiment, both the reasoning process and the final answer are examined to assess the model's performance and its potential to improve the explainability of LLM-generated results in the context of digital forensics.
The analysis focuses on how different parameter settings influence the quality, coherence, and reliability of the generated reasoning as well as the correctness of the resulting outputs.
We use quantitative metrics based on already existing metrics from the literature, as well as qualitative evaluation techniques.
The evaluation of the results was divided among three evaluators.

Since we are primarily interested in the reasoning process's potential to make the final answer comprehensible, we first evaluate the CoT.
For that, we decided to follow the four generic metrics presented in \cite{lee2025evaluating}: Factuality, Validity, Coherence, and Utility.

\begin{description}
    \item[Factuality] is evaluated using the number of factual errors in the reasoning process.
    A factual error is a fact taken either from the prompt or from incorrect common knowledge.
    An example of a factual error would be if the model refers to a specific date in the given timeline that is actually not included in the data.

    \item[Coherence] is evaluated similarly, but instead of the number of factual errors, the number of incoherences is computed. An incoherence is present when the model makes an inference while the preconditions for that inference are not present in the CoT.
    An example of an incoherence would be: ``The chat is about Breaking Bad'' without previously mentioning elements from the chat that relate to the TV show.

    \item[Validity] is evaluated using the number of logical/reasoning errors present in the reasoning process.
    A logical/reasoning error is a mistake in an inference that is due to a logical problem.
    An example of a logical error would be: ``We don't see a file copy event. The file was probably copied.''

    \item[Utility] is evaluated using the number of unnecessary steps, i.e., if it does not help the model to answer the question.
    An example of an unnecessary step is repeating the task without adding any new aspects to the following reasoning.
\end{description}

Besides that, we count the number of repetition loops in which the model gets stuck occasionally.
A repetition loop is an ongoing repetition of a word, a sequence of words, a sentence, or a sequence of sentences repeated twice or more.
In that case, the whole loop is considered a single step and is classified as one repetition.

We use these metrics to answer \hyperlink{rq1}{Research Question 1} and \hyperlink{rq2}{Research Question 2}.

To answer \hyperlink{rq3}{Research Question 3}, we evaluate the correctness of the final answer and the quality of the justification given for each answer.

\begin{description}
    \item[Correctness] is evaluated by comparing the predefined components of an answer for each experiment with the answer actually provided. 
    Each component is explicitly defined and must be correctly reflected in the response.
    For tasks such as suspicious message detection, only a single specific component is expected.
    For more complex tasks, the expected answers consist of two to eight components. Note that for the methodology generation, unrequested components present in the answer were penalized if incorrect.

    \item[] The quality of the \textbf{Justification} is evaluated qualitatively by the evaluator and converted to 1 or 0 (satisfying or not) to calculate a numerical value.
\end{description}

We compute separate scores for the CoT and the final answer, which we then use to compare the different experiments and parameters and to answer our research questions.
\begin{description}
    \item[] The \textbf{CoT Score} is the average of the normalized metric values introduced earlier.
    We normalized each CoT metric value by taking into account the number of steps in the reasoning process.
    These normalized values represent inverted metrics, as they indicate the relative proportion of steps that did not satisfy the metric (e.g., the relative number of factual errors is used to assess factuality).
    To obtain the expected metric, the normalized values are subtracted from 1: $1 - (\text{NB}{\text{elements}} / \text{NB}{\text{steps}})$.

    \item[] The \textbf{Final Answer Score} is a weighted combination of the correctness and justification values.
    Correctness, as described earlier, is normalized by the number of expected correct predefined answer components.
    The final answer score is a weighted mean, where correctness is counted three times and justification once.
\end{description}

Please note that the number of steps was determined manually by the evaluators, as one reasoning step can potentially comprise more than one sentence. The number of steps should not be confused with the number of tokens, which can be determined automatically.

%%%%%%%%%%%%%%%%%%%%%%%%
\section{Results}
\label{sec:results}
The findings highlight how variations in parameter settings affect both the model's reasoning behavior and the accuracy of its final outputs. Key trends and notable differences across experiments are summarized to provide a basis for the subsequent discussion.

Unfortunately, two (out of 60) experiments could not be analyzed as they reached the maximum number of 12500 tokens that we decided to authorize during the text generation\footnote{Both of them were generated using the deterministic method and had a high reasoning level setting.}.
They are excluded from the results analysis except stated otherwise.

\subsection{Quality of the Reasoning Process and the Final Answer}
This section discusses the quality of the reasoning process, the resulting CoT, and the quality of the final answer.

\begin{figure}[ht]
    \centering
    \includegraphics[scale=0.25]{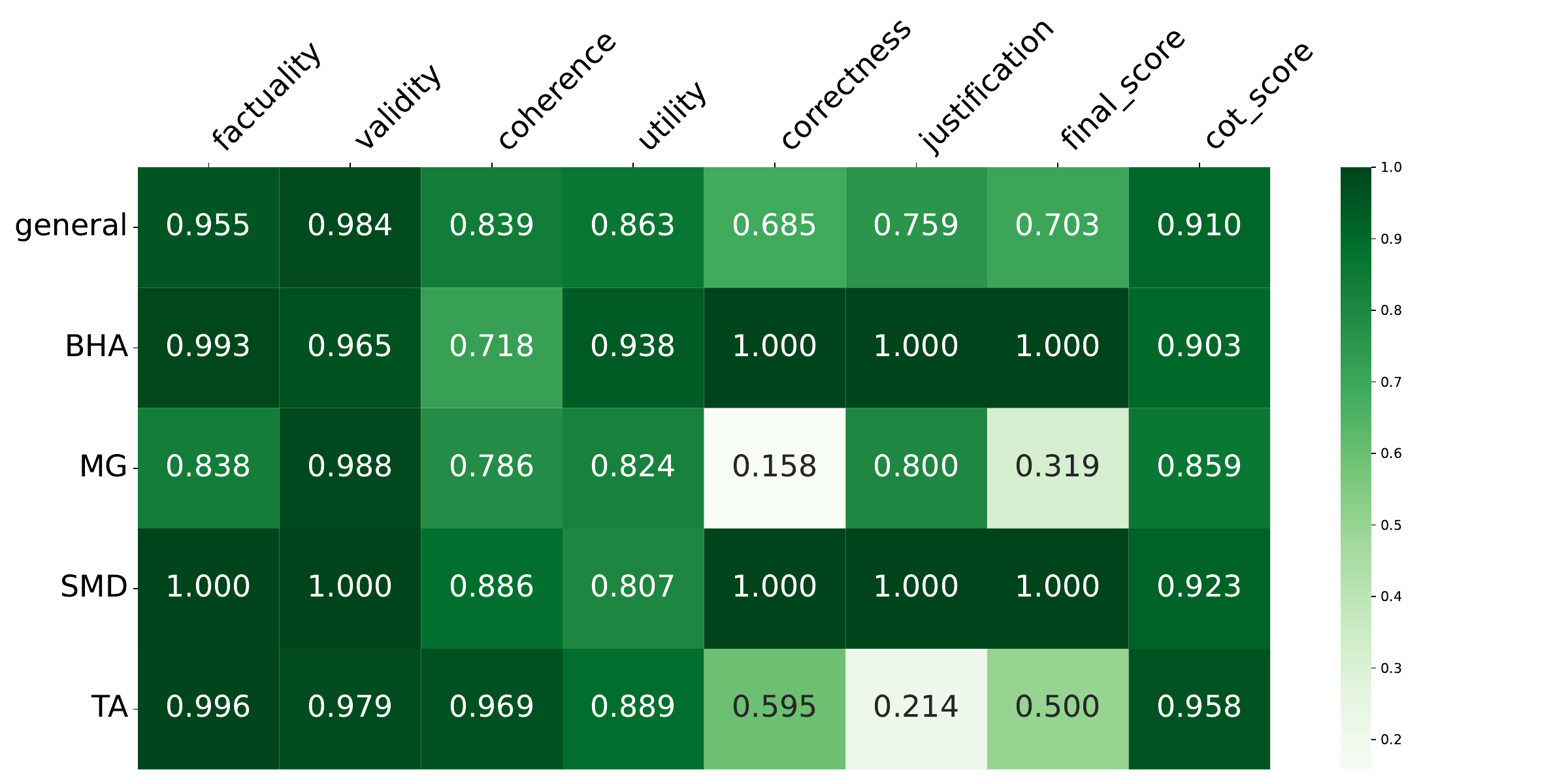}
    \caption{Averaged metric values for all experiments combined and for all experiments of each separate task. The `general' line represents the average of every metric value for all the samples. In contrast, each task represents the average of each metric value for the samples of that particular task. The closer a metric value is to one, the better it is.}
    \label{fig: fig3}
\end{figure}

Fig.~\ref{fig: fig3} shows the averaged metric values for all experiments combined and the averaged metric values for all experiments of each separate task.
Among the four tasks, the methodology generation and timeline analysis stand out for their lower correctness.
Both can be explained by the complexity of the task at hand.
In the case of methodology generation, the model often produced useless additional responses that were not asked for (for example, the model frequently included legal aspects in the response).
In the case of timeline analysis, the model had to choose between a yes or no answer despite the sparse data available.
In all cases, the model decided to go with the suspicion mentioned in the context, even though there was no clear evidence for it.
These two tasks also have the highest number of steps and tokens generated for the CoT, indicating a higher level of complexity.

The factuality is good overall, but lower for methodology generation.
This can be explained by the fact that the methodology generation had to use a lot of common digital forensics knowledge in the reasoning process, which was often inaccurate.
On the other hand, the three other tasks relied on the data provided in the prompt and therefore made fewer factual errors. 

Regarding the coherence, two outliers can be detected: the methodology generation and the log-file analysis.
For the former, incoherences were present when the model drafted the methodology structure by adding new steps or components that were not previously mentioned.
The latter often provided answers first and then explained them, which is not the correct reasoning order.

Exact repetitions were detected in texts generated for the methodology generation and suspicious message detection.

For the methodology generation, the final answer score was also generally better than the pure correctness value, as the final answer was often well justified, even if not very accurate.
On the other hand, answers were often not correctly justified for the timeline analysis, explaining why the final score is worse than the pure correctness value.

Therefore, \hyperlink{rq1}{Research Question 1} can be answered as follows: The quality of the generated CoTs differs among the different forensics tasks mainly due to their complexity. However, the CoT generally appears to be of good quality and useful.

\subsection{Impact of the Reasoning Level on the CoT Quality}
We now discuss the impact of the reasoning level on the quality of the generated CoT.

\begin{figure}[ht]
    \centering
    \includegraphics[scale=0.23]{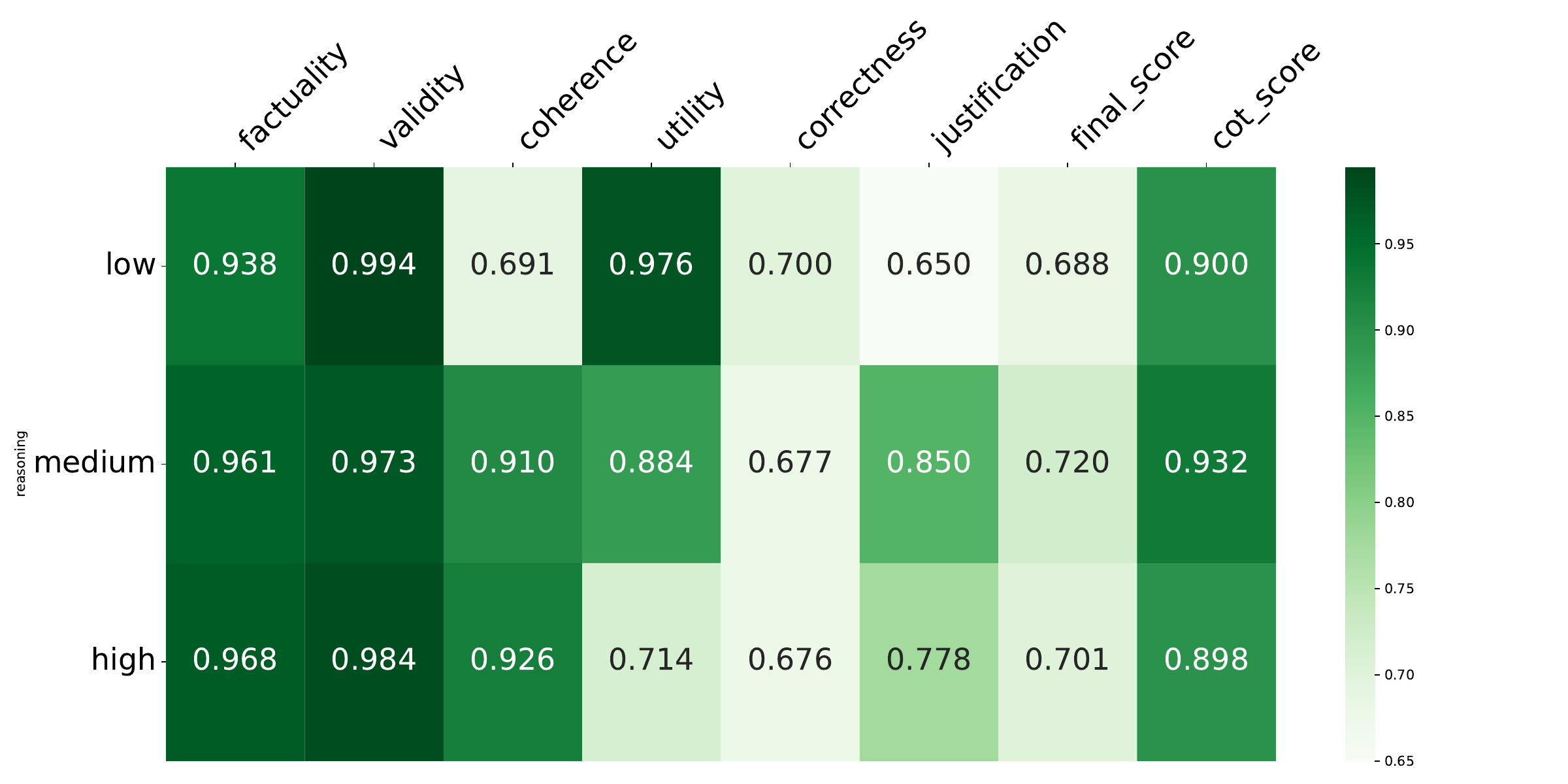}
    \caption{Mean of every metric value in general and for each task.}
    \label{fig: fig45}
\end{figure}

Fig.~\ref{fig: fig45} presents the average value of each metric across the different reasoning levels.
Each line represents the mean metric values for the samples generated at a given reasoning level.

Overall, the CoT score is highest at the medium reasoning level.
This can be explained by an increase in coherence and a decrease in utility as the reasoning level rises.
Higher reasoning levels produce longer CoTs with more steps and tokens.
When the number of steps is too small, the model fails to establish the necessary preconditions, which reduces coherence.
Conversely, when the number of steps becomes too large, the model tends to repeat content or explore irrelevant possibilities, thereby lowering utility.
The medium reasoning level appears to balance coherence and utility well.

Another interesting observation is that the number of repetition loops scales with the reasoning level: the higher the reasoning level, the greater the number of repetition loops.
It is worth noting that repetition loops were only observed at the medium-level reasoning (in suspicious message detection) and at the high-level reasoning (in both suspicious message detection and methodology generation).

Additional metric values were computed for each task, revealing that the coherence trend differs between methodology generation and timeline analysis.
Both tasks achieve the highest coherence at the medium reasoning level.
This can be explained by the timeline analysis task's tendency to produce overly sophisticated hypotheses not grounded in the input data, and by the generation of a larger number of methodology drafts at the medium and high reasoning levels, elements of the reasoning process that exhibited the greatest incoherence.

Therefore, \hyperlink{rq2}{Research Question 2} can be answered as follows: The reasoning level directly impacts the quality of the CoT, with the medium level of reasoning resulting in the best quality.

\subsection{Impact of the Reasoning Process on the Final Answer}
We also evaluated the impact of the reasoning process (the model's ability to `think') on the quality of the final answer.

\begin{figure}[ht]
    \centering
    \includegraphics[scale=0.55]{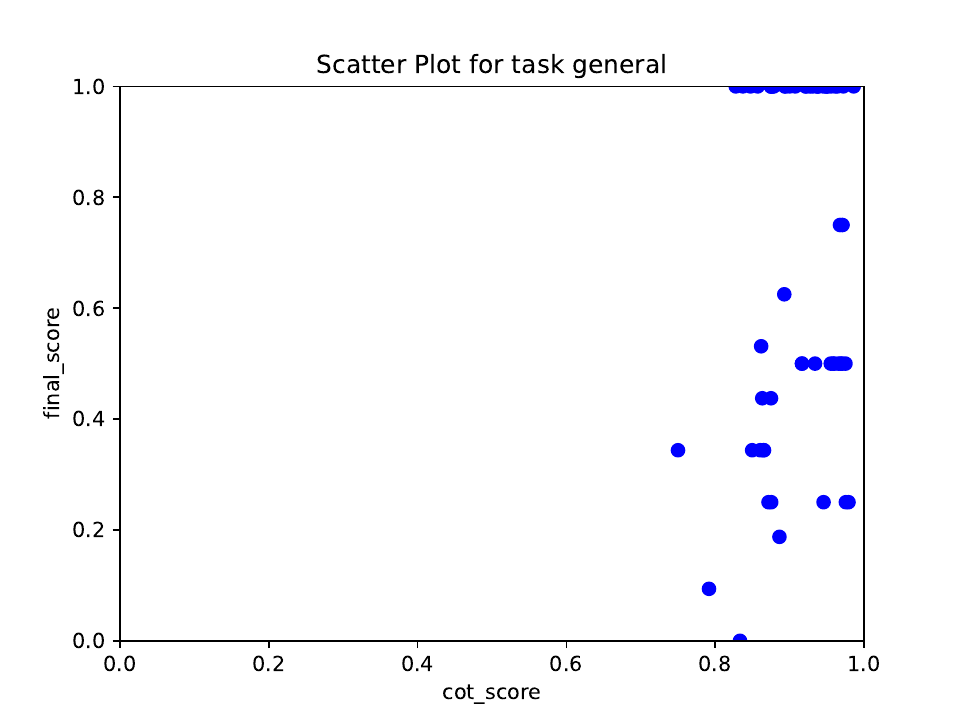}
    \caption{Scatter plot showing the score of the reasoning process (CoT) on the x-axis and the score of the final answer (final) on the y-axis for all experiments.}
    \label{fig: fig2}
\end{figure}

When plotting the CoT score against the final answer scores, no clear trend or observable effect can be identified, as shown in Fig.~\ref{fig: fig2}.
The plot shows the score obtained for the reasoning process on the x-axis and the score obtained for the final answer on the y-axis for all experiments.
The higher the score, the better the evaluated element.
This absence of a trend or relation could be interpreted as the absence of an impact of the reasoning process on the quality of the final answer.
However, the quality of the final answer is also influenced by the complexity of the task and the difference between the number of steps or elements measured in the reasoning process and in the final answer.
For example, for some of the tasks, the reasoning process at higher levels of reasoning involves significantly more steps or elements than appear in the final answer.

We can therefore answer \hyperlink{rq3}{Research Question 3} by: No, the quality of the CoT does not directly impact the quality of the final answer.

\subsection{Impact of the Generation Strategy}
During the inferences, two types of generation were used: sampling (with a temperature of 0.7) and a deterministic approach (similar to a temperature of 0). Most of the exact repetitions were observed in the texts generated using the deterministic approach. 

These texts also had a lower final score, a higher CoT score, a smaller number of reasoning steps, and a higher number of tokens generated for the reasoning process. This could be explained by the higher number of repetitions, which increases the number of tokens in the CoT. The behavior was different for timeline analysis and log-file analysis, two tasks without repetition.

To provide an overview of the overall results, we have listed them in detail in Tab.~\ref{tab:results}.

\begin{table*}[h]
\resizebox{\textwidth}{!}{%
% \begin{tabular}{@{}lll@{}}
\begin{tabular}{l|l|l|l|l|l|l|l|l|l|l|l}
\toprule
% 1, SMD, reasoning, temperature, sampling
\textbf{ID} & \textbf{Task} & \textbf{Reasoning} & \textbf{Sampling} & \textbf{Factuality} & \textbf{Coherence} & \textbf{Validity} & \textbf{Utility} & \textbf{CoT Score} & \textbf{Correctness} & \textbf{Justification} & \textbf{Answer Score} \\ \midrule
Mean & General & - & - & 0.955 & 0.984 & 08.39 & 0.863 & 0.910 & 0.685 & 0.759 & 0.703 \\
\midrule
1-4 & SMD & Low & Yes & 1.000 & 1.000 & 0.775 & 1.000 & 0.944 & 1.000 & 1.000 & 1.000 \\
5 & SMD & Low & No & 1.000 & 1.000 & 0.800 & 1.000 & 0.950 & 1.000 & 1.000 & 1.000 \\
6-9 & SMD & Medium & Yes & 1.000 & 1.000 & 0.934 & 0.920 & 0.964 & 1.000 & 1.000 & 1.000 \\
10 & SMD & Medium & No & 1.000 & 1.000 & 0.944 & 1.000 & 0.986 & 1.000 & 1.000 & 1.000 \\
11-14 & SMD & High & Yes & 1.000 & 1.000 & 0.936 & 0.454 & 0.847 & 1.000 & 1.000 & 1.000 \\
15 & SMD & High & No & 1.000 & 1.000 & 0.957 & 0.617 & 0.894 & 1.000 & 1.000 & 1.000 \\
\midrule
Mean & SMD & - & - & 1.000 & 1.000 & 0.886 & 0.807 & 0.923 & 1.000 & 1.000 & 1.000 \\
\midrule
16–19 & BHA & Low & Yes & 1.000 & 1.000 & 0.482 & 1.000 & 0.871 & 1.000 & 1.000 & 1.000 \\
20 & BHA & Low & No & 1.000 & 1.000 & 0.600 & 1.000 & 0.900 & 1.000 & 1.000 & 1.000 \\
21–24 & BHA & Medium & Yes & 0.991 & 0.929 & 0.773 & 0.929 & 0.906 & 1.000 & 1.000 & 1.000 \\
25 & BHA & Medium & No & 1.000 & 0.931 & 0.793 & 1.000 & 0.931 & 1.000 & 1.000 & 1.000 \\
26–29 & BHA & High & Yes & 0.985 & 0.967 & 0.908 & 0.853 & 0.928 & 1.000 & 1.000 & 1.000 \\
30 & BHA & High & No & - & - & - & - & - & - & - & - \\
\midrule
Mean & BHA & - & - & 0.993 & 0.965 & 0.718 & 0.938 & 0.903 & 1.000 & 1.000 & 1.000 \\
\midrule
31–34 & MG & Low & Yes & 0.789 & 1.000 & 0.520 & 1.000 & 0.827 & 0.250 & 0.75 & 0.375 \\
35 & MG & Low & No & 0.667 & 1.000 & 0.667 & 1.000 & 0.833 & 0.000 & 0.000 & 0.000 \\
36–39 & MG & Medium & Yes & 0.837 & 0.982 & 0.935 & 0.704 & 0.865 & 0.219 & 0.750 & 0.352 \\
40 & MG & Medium & No & 0.913 & 1.000 & 0.913 & 0.957 & 0.946 & 0.000 & 1.000 & 0.250 \\
41–44 & MG & High & Yes & 0.899 & 0.976 & 0.867 & 0.736 & 0.869 & 0.063 & 1.000 & 0.297 \\
45 & MG & High & No & 0.893 & 0.987 & 0.927 & 0.645 & 0.863 & 0.250 & 1.000 & 0.438 \\
\midrule
Mean & MG & - & - & 0.838 & 0.988 & 0.786 & 0.824 & 0.859 & 0.158 & 0.800 & 0.319 \\
\midrule
46–49 & TA & Low & Yes & 0.985 & 0.972 & 0.944 & 0.881 & 0.946 & 0.583 & 0.000 & 0.438 \\
50 & TA & Low & No & 1.000 & 1.000 & 0.875 & 1.000 & 0.969 & 0.667 & 0.000 & 0.500 \\
51–54 & TA & Medium & Yes & 1.000 & 0.972 & 0.994 & 0.914 & 0.970 & 0.583 & 0.500 & 0.563 \\
55 & TA & Medium & No & 1.000 & 0.988 & 0.996 & 0.852 & 0.959 & 0.333 & 1.000 & 0.500 \\
56–59 & TA & High & Yes & - & - & - & - & - & - & - & - \\
60 & TA & High & No & 0.999 & 0.986 & 0.985 & 0.854 & 0.956 & 0.667 & 0.000 & 0.500 \\
\midrule
Mean & TA & - & - & 0.996 & 0.979 & 0.969 & 0.889 & 0.958 & 0.595 & 0.214 & 0.500 \\ \bottomrule
\end{tabular}%
}
\caption{Overall metric values for all experiments, including the average values for all experiments and per task.}
\label{tab:results}
\end{table*}

\subsection{Qualitative Results}
Finally, we discuss the qualitative evaluation results for each of the four digital forensics test scenarios.

\subsubsection{Suspicious Message Detection}
In the context of the suspicious message detection task, the reasoning patterns across samples show notable consistency.
Samples usually start in a similar way: by assuming that the chat refers to the Breaking Bad TV show and then evaluating its relevance. The final answers are consistently correct and well justified.

CoTs produced with low reasoning levels, while easier to read and follow, do not provide sufficient explanation for the final answer. In contrast, the medium and high reasoning levels allow for more extensive reasoning and offer better justification, though both also contain more repetitive content.

\subsubsection{Bash History Analysis}

In the server attack investigation task context, the model’s reasoning process follows a recurring pattern.
The model often provides an initial answer to the question at the very beginning of the reasoning process, relying on only a few elements.
It then proceeds to expand the evidence list or explore alternative hypotheses.
The reasoning sequence is often repetitive: the model returns to the question, answers it, introduces additional evidence or tests new hypotheses, and then revisits the question.

References to the commands included in the input data are generally very accurate.
However, the model occasionally has difficulties with the attack flow; for instance, it confuses a reverse shell with a shell.php file downloaded from the targeted server.
Despite these issues, the final answers are consistently correct and well justified.

CoTs produced at the low reasoning level are insufficient to explain the final answer, as the response is given directly without any reasoning or supporting evidence.
The medium reasoning level introduces more elements but may not fully justify the final answer.
In contrast, the high reasoning level is likely sufficient to explain all aspects, although it contains considerably more repetitive content.

\subsubsection{Methodology Generation}
The reasoning process generated during the methodology generation task differs slightly from the others. At the beginning, the model reiterates and elaborates on what needs to be done, but soon shifts into producing a list of elements to be included in the final answer.

Samples generated with the high reasoning level, and one of the medium-level samples, also attempt to draft one or several possible structures for the final answer during the reasoning process.

The low reasoning level is insufficient to justify the final answer. In contrast, the medium and high levels tend to repeat content extensively without adding meaningful value. This effect is particularly pronounced at the high reasoning level.

Both the medium and high reasoning levels frequently mention data sources such as database or log file names and file paths. While potentially interesting, these details were not requested and are often inaccurate. The model also exhibits difficulties with tool identification, frequently combining the names of well-known tools or assigning them to inappropriate tasks (e.g., Cellebrite Physical Analyzer is often cited as an acquisition tool, although this role should be attributed to Cellebrite 4PC). The mention of a write-blocker is also common during acquisition steps; while required for certain types of storage, it cannot be practically used with smartphones.

The final answers often include numerous elements and detailed sub-steps that were not explicitly requested. This behavior may stem from the prompt's instruction to provide a `detailed methodology'.

\subsubsection{Timeline Analysis}

The timeline analysis represents the most complex task. The input data are sparse and incomplete, and the available event information is limited due to context size constraints. There is no definitive evidence that the file in question was copied to the USB drive or that the Internet connection was deactivated.
However, the model must provide a definitive yes-or-no answer, as specified in the prompt. Consequently, it struggles with the ambiguity between the data and the task, often reasoning along the lines of: ``We need to answer yes or no. We could say `no' because we cannot confirm that the Internet was deactivated, but we could also say `yes' because there is no evidence of Internet usage.''

This binary requirement prevents the model from expressing uncertainty in its final answer, even though such uncertainty is explicitly discussed during the reasoning process. As a result, certain statements may appear definitive in the final answer while being presented as uncertain in the chain of thought. In some cases, the model even arrives at an incorrect conclusion despite recognizing that the evidence does not clearly support it. In one instance, the model went so far as to question whether parts of the timeline might originate from another device.

For this task, access to the reasoning section of the generated text is particularly important, as it reveals the model's difficulty concluding and highlights its awareness of uncertainty.

The model also reflects on user assumptions regarding the Internet being deactivated and the type of answer expected by the user. At higher reasoning levels, it begins to explore unsupported and speculative hypotheses, such as the already mentioned hypothesis that the timeline belongs to another machine, the presence of a local DNS server resolving URLs to local IPs, the use of proxies, or missing data entries.

Therefore, \hyperlink{rq4}{Research Question 4} can be answered as follows: Yes, the reasoning process and the generated CoT can be used to improve the explainability of LLM-generated answers to specific digital forensics tasks, especially in cases where the task includes aspects of ambiguity or if the input data is not complete.

%%%%%%%%%%%%%%%%%%%%%%%%
\section{Discussion}\label{sec:discussion}

\subsection{Discussion of the Results}
\paragraph{Quality of the CoT} 
The results obtained for the chosen metrics indicate good performance for the model's CoT, with an average value of the four metrics systematically over 0.859. In the same way, the qualitative analysis indicates that many final answers can be understood/justified using the content of the CoTs generated with a medium or high reasoning level.

\paragraph{Quality of the Final Answer} 
The quality of the final answer varies significantly, with a lower value for the two tasks that were considered the most complex: the methodology generation and timeline analysis. This seems to indicate that the quality of the final answer is reduced if the input data and request provided in the prompt are complex (timeline analysis) or when the task's success relies on digital forensics common knowledge (methodology generation). Therefore, the final output is likely to be of higher quality for straightforward tasks that depend primarily on the data provided directly in the prompt.

\paragraph{Impact of the Reasoning Level} 
Regarding the impact of the reasoning level, the quantitative and qualitative analyses both indicate that the medium reasoning level performs better. While the reasoning level does not significantly impact the factuality and validity, there is a clear relation between the reasoning level, the coherence, and the utility. With a higher reasoning level, the model has more `room' to present the preconditions before making inferences, but it also tends to repeat more content. Therefore, the medium level represents a good balance between coherence and utility.
As mentioned previously, there is an exception to the coherence between methodology generation and timeline analysis, where the medium level is better than the high level.

\paragraph{Impact of the CoT on the Final Answer} 
The quality of the final answer seems to be impacted by the task's difficulty, not by the quality of the CoT. No clear impact of the CoT quality on the final answer quality could be identified. This means that a high-quality CoT could lead to a low-quality final answer, and that a final answer could be of high quality even if the CoT is of low quality.

\paragraph{Utility of the CoT} 
Based on the obtained results, we consider the model's CoT a useful tool for better understanding how the model arrived at its answer.
An investigator could, for example, quickly understand why the model determined that an attack took place by examining the reasoning process and reading the explanations, which include references to logs present in the bash history. In scenarios where the request is ambiguous (such as the yes/no request for the timeline analysis), the model shows uncertainty in the CoT while being confident in the final answer.
Still, it is insufficient to actually explain how the model generated the answer (or the CoT) and does not constitute explainable AI.
Forgetting this could lead investigators to model over-trust. In such a case, the CoT's positive impact would turn into a factor favoring errors.
Therefore, the reasoning process is considered a useful component, but it is not considered sufficient to explain how the model generated the answer. At the end of the day, an RLM remains a language model that is trained to predict the most probable next tokens in a given context. It should therefore be regarded as a tool whose statements must always be validated. This is a process in which CoT can be of assistance.

%%%%%%%%%%%%%%%%%%%%%%%%
\section{Limitations and Next Steps}\label{sec:limitations}
The main limitations of this study can be classified into three categories:
\begin{description}
    \item[Study's Scale:] The study considers a single model that is limited in size. The reason for this choice was the exploratory nature of the study and the search for a balanced size model, i.e., a model that is not too small and that could fit on most of the non-professional GPUs to better mimic operational constraints. 
    Testing more models of various sizes should be investigated in future research. 
    Moreover, due to the manual nature of the evaluation, only 60 samples (including two that were discarded) distributed over four tasks were evaluated, reducing the generalizability of the results.
    \item[Metrics' Choice:] While factuality, validity, and utility were easy to understand and evaluate, coherence was a complex metric. Determining if all the preconditions are met before making the inference is complex, especially in the CoT, with preconditions scattered throughout the previous steps.
    This study considered four general metrics. In future work, more metrics, including automatically computed metrics, could be considered.
    \item[Evaluation Process:] Although metrics were defined to reduce subjectivity during the evaluation process, the three evaluators likely did not evaluate the results in the exact same way. For example, the division of the CoT into reasoning steps could have been done differently by the evaluators, or a reasoning step could be considered normal for one evaluator and incoherent for another.
    Finally, although it helped reduce the impact of the number of steps, the normalization process did not completely remove the bias introduced by the significant size difference between the reasoning processes generated by different levels of reasoning.
\end{description}

%%%%%%%%%%%%%%%%%%%%%%%%
\section{Conclusion}\label{sec:conclusion}
The current LLMs' lack of explainability reduces the compatibility of such models with the digital forensics requirements. To advance the goal of explainable large language models, we explored the use of the chain of thought mechanism implemented in gpt-oss-20b. This model represents a new class of language models specifically optimized for reasoning tasks, designed to generate intermediate reasoning steps before producing the final answer presented to the user. We selected four common forensic tasks and related prompts. We then provided these prompts to the model, varying the task, the reasoning level, and the generation strategy (sampling activated vs. sampling deactivated). We manually evaluated the reasoning process using four metrics: factuality, validity, coherence, and utility. The final answers were then evaluated based on the correctness and justification provided by the model. Comments from the evaluators were also compiled and presented.

\emph{What is the quality of the CoT generated by gpt-oss for a given digital forensics task?} In general, the CoT evaluated was of quality, based on the four evaluated metrics. The methodology and timeline analysis had the worst results for the final answer, indicating higher complexity tasks.

\emph{How does the adjustable level of reasoning impact the quality of the CoT?} The factuality and validity generally remain stable between the reasoning levels. On the other hand, the coherence and utility generally tend to increase and decrease when the reasoning levels increase. The most balanced level is the medium reasoning level.

\emph{Does the quality of the CoT impact the quality of the final answer?} No, the impact of the CoT's quality on the final answer's quality could not be found. The quality of the final answer seems to be driven mainly by the difficulty of the task.

\emph{Can the CoT effectively assist in understanding and explaining why the model reached a certain conclusion?} Yes, the CoT can help understand or justify the generation of the final answer, but it is insufficient to explain it and does not constitute a form of explainable AI.

Future research should investigate different reasoning models, consider additional digital forensics tasks, and use a combination of manually and automatically computed metrics to reduce the subjectivity introduced by manual evaluation. It should also evaluate more samples per task to increase the generalizability of the results.

Although the results are encouraging, it is not recommended to use CoT reasoning as an explanation for the final answer, especially if the results are presented in court. However, CoT can be considered as an aid to understanding and verifying the final answer.

%%%%%%%%%%%%%%%%%%%%%%%%
%\section*{Acknowledgments} \textit{blinded for review}

\section*{Disclosure of AI-Assisted Writing Tools}
Several authors used ChatGPT and Grammarly to help with tasks such as revising, condensing text, and addressing grammatical errors, typos, and awkward phrasing. All AI-generated suggestions were thoroughly reviewed and adjusted where needed to ensure they accurately reflected the authors’ intended meaning before being incorporated into the paper.

%%%%%%%%%%%%%%%%%%%%%%%%
\section*{CRediT Authorship Contribution Statement}

%Conceptualization, Methodology, Investigation, Writing - Original Draft, Writing - Review \& Editing, Visualization, supervision.

\textbf{Gaëtan Michelet:} Conceptualization, Methodology, Software, Investigation,  Writing - Original Draft,  Visualization.
\textbf{Janine Schneider:} Methodology, Investigation, Writing - Original Draft.
\textbf{Aruna Withanage:} Investigations.
\textbf{Frank Breitinger:} Conceptualization, Writing - Review \& Editing, Supervision.

%%%%%%%%%%%%%%%%%%%%%%%%
\section*{Declaration of Interest}
The authors declare that they have no known competing financial interests or personal relationships that could have appeared to influence the work reported in this paper.


%\bibliographystyle{elsarticle-num-names}
%\bibliographystyle{unsrtnat}
\bibliographystyle{bibl/model5-names}
\bibliography{bibl/literature.bib}

%% The provided style fileelsarticle-num.bst formats references in the required Procedia style





% \end{thebibliography}

\appendix
\clearpage

\section{Appendix - Digital Forensics Tasks}
\label{sec:appendix}

In this section, we list the details of the digital forensic tasks that we have instructed the RLM to solve.
The tasks are sorted by complexity.

\subsection{Suspicious Message Detection (SMD)}

\begin{description}
    \item [Description:] The model is provided with information about a case investigated and messages extracted from a chat conversation. It must then determine whether the chat is relevant to the investigation or not.
    \item [Justification:] This task is time-consuming and very common during digital forensic investigations. A reasoning process is required to successfully determine if the chat is relevant to the investigation.
    \item[Context and instruction:]  The context in this scenario is a drug-related investigation in which Alex, a fictional character, is suspected of dealing. A long chat conversation involving Alex was secured and is provided as input data. The model is instructed to determine whether the chat is relevant to the ongoing investigation or not. The answer needs to be justified.
    \item [Input data:] The chat was generated using GPT-5 mini. The discussion revolves around the Breaking Bad TV show, with the two participants commenting on implications related to the show. In the initial chat, direct mentions of the show's name were present and manually removed to complicate the task. The text was manually adjusted to remain coherent without mentioning the show. Note that the names of certain characters from the show were not removed.
    \item [Expected answer:] The model is supposed to indicate that the chat is irrelevant for the investigation and provide a justification.
\end{description}

\subsection{Bash History Analysis (BHA)}

\begin{description}
    \item [Description:] The model is provided with the bash history extracted from the laptop belonging to someone suspected of trying to attack a server. It must determine if an attack was undertaken from that laptop and, if so, who was the target.
    \item [Justification:] This task is time-consuming and requires reasoning to determine whether some commands are potentially harmful and who could be targeted. 
    % The bash history is a completely textual log file.
    \item[Context and instruction:] This scenario involves investigating a person suspected of possibly attempting a server attack. The computer was seized, the disk was imaged, and the bash history was extracted from the seized Linux system. The model is tasked to determine if an attack was undertaken from the computer, and if so, who the target of that attack was. It must also justify the answers.
    \item [Input data:] The input data is a bash history generated using GPT-5 mini. After the initial generation, the bash history was manually reviewed to remove repetitive elements and rename overly obvious test and example file and website names. In addition, some of the commands were adjusted.
    \item [Expected answer:] The model should acknowledge that an attack could have been undertaken from that computer, that the server with IP 100.200.30.40 was the target, and justify these answers.
\end{description}

\subsection{Methodology Generation (MG)}

\begin{description}
    \item [Description:] The model receives information about a case under investigation and the associated investigative questions. It must then design a methodology that can be used to answer these questions.
    \item [Justification:] This is a prevalent task during investigations where a methodology must be established according to the questions asked by the investigator. The creation of such a targeted methodology requires reasoning.
    \item[Context and instruction:] The scenario centers on an investigation related to a murder. The suspect's Samsung S23 was seized and sent to the forensic lab for analysis. Investigators are tasked with determining the suspect's location on January 15, 2024, following the NIST four-phase framework (Collection, Examination, Analysis, Reporting). These two elements serve as the input data. Based on the NIST four-phase framework, the model is then required to outline a methodology that enables investigators to answer the question regarding the suspect's whereabouts on the day of the murder. For each step, the model should also justify and recommend appropriate tools to carry out the procedures.
    \item [Input data:] For this task, the input data consists of the question posed by the investigators: ``Can you determine the suspect's location on January 15, 2024, the day of the murder?'' and the investigation model employed by the institution (the NIST four-phase framework).
    \item [Expected answer:] The model must provide the draft of a methodology following the NIST four-phase framework that can help determine the suspect's location on the date of interest. Each step must be justified, and tools should be suggested whenever possible.
\end{description}

\subsection{Timeline Analysis (TA)}

\begin{description}
    \item [Description:] The model is provided with information about a suspected data theft and a timeline obtained from a forensic image. It must then determine whether the timeline data can confirm the theft.
    \item [Justification:] This task is particularly time-consuming. Moreover, intense reasoning is required in order to be undertaken. 
    % Finally, the timeline can easily be converted to CSV, making it a text-based input data. 
    This makes timeline analysis a good fit for this experiment.
    \item[Context and instruction:] The context in this scenario is an investigation requested by a company named `Out-of-Date GmbH'. This company assumes that one of its employees stole its world-famous secret recipe. They assume the employee got access to a mainframe server (which is not connected to the Internet) and copied the secret recipe to another device. They think it probably happened during maintenance on 04/19/2017. A timeline of the mainframe server filtered on the date of interest is provided as input data. The model is tasked to determine if something suspicious happened, and if so, what? It must also determine if the Internet connection was actually deactivated and justify all the answers.
    \item [Input data:] The input data is a modified version of a timeline manually generated for an exercise on digital forensics in 2017. Given the gpt-oss context window and GPU memory limit, we decided to remove some irrelevant events and fields from that timeline in order to reduce the number of tokens in the prompt provided to the model. The events were first filtered based on the date of interest (04/19/2017). Afterward, Eric Zimmerman's timeline explorer was used to filter out events that were not highlighted by the tool. Highlighted and therefore essential events are: file opening, web history, deleted data, execution, device or USB usage, folder opening, and log file.
    \item [Expected answer:] The model must answer that a USB drive has been plugged into the machine, that the secret recipe can indeed be found on the machine, and that the data is insufficient to prove that the recipe has been stolen. Additionally, it must indicate that the Internet connection was likely activated at some point and justify all these answers. The challenge with this task is that the data is actually insufficient to answer the investigative questions clearly. Nevertheless, the model is asked to answer with yes or no. This is intended to simulate a situation in which a user forces the model to decide on an answer even though the available data is actually insufficient for this.
\end{description}


\end{document}